\documentclass[twocolumn]{aastex631}

\usepackage{breqn}

\received{April 4, 2024}
\revised{July 21, 2024}
\accepted{August 16, 2024}
\published{October 10, 2024}
\submitjournal{The Astrophysical Journal}

\shorttitle{Transit Time Biasing by Limb Asymmetry}
\shortauthors{Murphy, Beatty, \& Apai}

\begin{document}

\title{An Analytic Characterization of the Limb Asymmetry - Transit Time Degeneracy}

\correspondingauthor{Matthew M. Murphy}
\email{mmmurphy@arizona.edu}

\author[0000-0002-8517-8857]{Matthew M. Murphy}
\affiliation{Steward Observatory, 933 N. Cherry Avenue, Tucson, AZ, 85719, USA}

\author[0000-0002-9539-4203]{Thomas G. Beatty}
\affiliation{Department of Astronomy, University of Wisconsin--Madison, 475 N. Charter Street, Madison, WI, 53706, USA}

\author[0000-0003-3714-5855]{D\'aniel Apai}
\affiliation{Steward Observatory, 933 N. Cherry Avenue, Tucson, AZ, 85719, USA}
\affiliation{Lunar and Planetary Laboratory, University of Arizona, Tucson, AZ, 85705, USA}

\begin{abstract}
Atmospheres are not spatially homogeneous. This is particularly true for hot, tidally locked exoplanets with large day-to-night temperature variations, which can yield significant differences between the morning and evening terminators -- known as limb asymmetry. Current transit observations with the James Webb Space Telescope (JWST) are precise enough to disentangle the separate contributions of these morning and evening limbs to the overall transmission spectrum in certain circumstances. However, the signature of limb asymmetry in a transit light curve is highly degenerate with uncertainty in the planet's time of conjunction. This raises the question of how precisely transit times must be measured to enable accurate studies of limb asymmetry, in particular with JWST. Although this degeneracy has been discussed in the literature, a general description of it has not been presented. In this work, we show how this degeneracy results from apparent changes in the transit contact times when the planetary disk has asymmetric limb sizes. We derive a general formula relating the magnitude of limb asymmetry to the amount by which it would cause the apparent time of conjunction to vary, which can reach tens of seconds. Comparing our formula to simulated observations, we find that numerical fitting techniques add additional bias to the measured time, of generally less than a second, resulting from the occultation geometry. We also derive an analytical formula for this extra numerical bias. These formulae can be applied to planning new observations or interpreting literature measurements, and we show examples for commonly studied exoplanets. 
\end{abstract}


\section{Introduction} \label{sec:intro}
Atmospheres are not perfectly uniform. Observational surveys of brown dwarfs and planetary mass objects have found robust evidence that their atmospheres are heterogeneous and variable \citep[e.g.,][]{buenzli2014browndwarfs, metchev2015browndwarfs,apai2017}. For transiting exoplanets specifically, there is abundant evidence for day-to-night variation in atmospheric temperature and cloudiness \citep[e.g.,][among many others]{beatty2019daynighttemperatures, may2022hotjupitertrends} driven by the contrast in irradiation received by the day and night hemispheres. Three-dimensional atmospheric circulation models also predict that this day-to-night contrast will drive differences between the morning and evening terminator regions \citep[e.g.,][]{kataria2016_gcmgrid, line2016, powell2019}, which are known as ``limb asymmetries" because these terminators form the limbs of the planetary disk when seen in transit. 

Observations are beginning to confirm these predictions of morning-to-evening limb asymmetry. To date, most measurements of limb asymmetry have come from high spectral resolution ground-based spectroscopy, which can probe for wind speed and/or chemical abundance gradients through variable Doppler shifting between different periods of the planet's transit \citep{ehrenreich2020_wasp76b, bourrier2020_wasp121b, hoeijmakers2020_mascara2b, kesseli2021_wasp76b, borsa2021_wasp121b}. These observations have focused on ultra-hot exoplanets ($\gtrsim$2100~K), on which atomic metals like Li, Fe, and Mn (see \cite{kesseli2021_wasp76b} for an overview) are expected to be abundant and have numerous absorption lines that can only be probed at high spectral resolution. For example, \cite{ehrenreich2020_wasp76b} find strong, blue-shifted iron absorption during the transit egress of WASP-76b that is not present during ingress, suggesting that gaseous iron is present in the planet's evening terminator but not the morning terminator, implying condensation or rain-out on the cool nightside. 

The James Webb Space Telescope (JWST) has proven capable of directly detecting limb asymmetry at low spectral resolution as well. The observational signature of limb asymmetry in JWST transit observations are apparent time of conjunction variations with wavelength \citep{feinstein2022wasp39, rustamkulov2022ers} and changes in the chromatic transit depth between ingress and egress \citep{murphy:inreview, espinoza:inprep, delisle:inprep}. One advantage of JWST is that it captures the transmission spectra, both overall and of each limb's contribution, over a much wider wavelength range than is generally possible in high-resolution spectroscopy from the ground. This enables more detailed retrievals of the atmospheric temperature--pressure structure, molecular abundances, and aerosol properties in each spatial region using JWST. Extracting these separate contributions is vital to studying and understanding the circulation of heat and molecules within atmospheres. Also, previous analyses have shown that, if such differences are ignored, the properties retrieved from observed data can be biased significantly from the truth \citep{feng2016, caldas2019, taylor2020}. 

One major obstacle to measuring limb asymmetry via transmission spectroscopy is a degeneracy between the effect of limb asymmetry on the planet's transit light curve and uncertainty in the planet's time of conjunction. This degeneracy exists because the observable manifestation of limb asymmetry is a wavelength-dependent difference in the apparent radius of the morning and evening limbs. Not only does this alter the depth and shape of the light curve during ingress or egress, it also causes the transit to start and end slightly earlier or later. These changes in timing can be approximated by varying the planet's time of conjunction -- translating the entire transit slightly earlier or later in time. For example, numerical simulations find that measuring limb asymmetry of approximately a scale height difference in radius can require knowing the time of conjunction to within less than one second \citep{line2016, Espinoza2021_catwoman2}. While this degeneracy has been introduced and its magnitude discussed several times in the literature \citep[e.g. see][]{vonparis2016, line2016, powell2019, Espinoza2021_catwoman2}, a generally applicable analytical description of the degeneracy does not yet exist. Here, we derive an analytical description of the degeneracy between limb asymmetry and transit timing, and compare it to commonly used numerical techniques. This can be used for planning and interpreting observations designed to investigate limb asymmetry. 

\section{Analytical Method} \label{sec:analytical}
    \subsection{Defining the Problem} \label{subsec:analytical_motivation}

\begin{figure*}
    \centering
    \includegraphics[width=\textwidth]{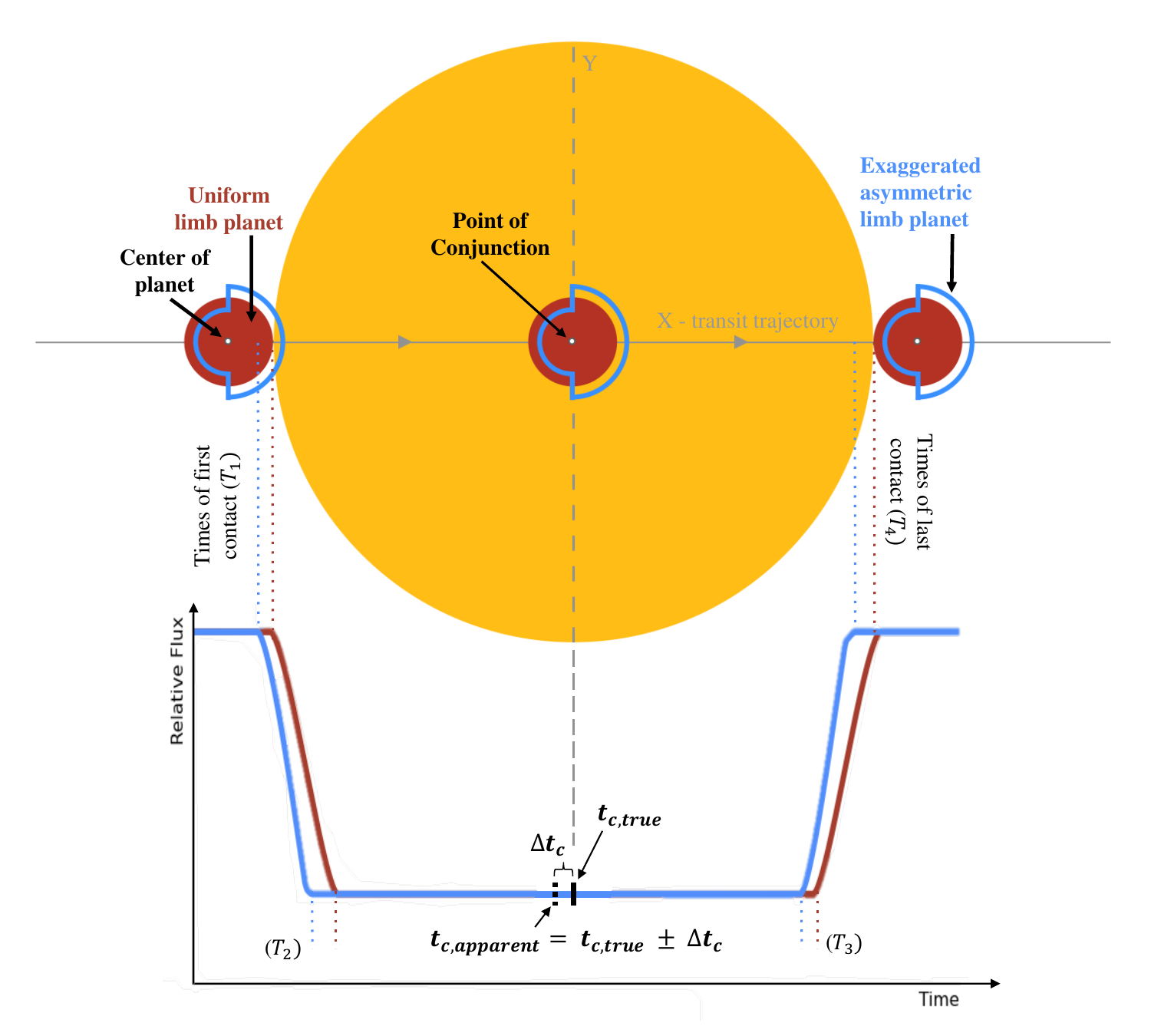}
    \caption{Exaggerated schematic illustrating the difference in the transit light curves of an asymmetric-limb planet and a uniform-limb planet. Limb darkening is omitted for visual simplicity. Both planets share the same time of conjunction. Due to the difference in their leading and trailing limb radii though, the transit contact times of each planet differ. An observer that measures the light curve of the asymmetric-limb planet, which is shown in blue, and fits their data using a uniform-limb transit model would infer a biased time of conjunction because these changes in the contact times effectively shift the transit earlier or later in time. This is the origin of the degeneracy between limb asymmetry and variations or uncertainty in a planet's time of conjunction.
    }
    \label{fig:exaggerated_intro_schematic}
\end{figure*}

Consider an example planet with asymmetric limbs, where each limb is represented by a semicircle joined to the other along its flat edge \citep[as in Figure~\ref{fig:exaggerated_intro_schematic}, inspired by][]{Espinoza2021_catwoman2, catwoman1}. The evening limb has radius $R_{\mathrm{evening}}$ and the morning limb radius differs by some amount $\Delta R$, so that $R_{\mathrm{morning}} = R_{\mathrm{evening}} + \Delta R$. This $\Delta R$ can be positive or negative depending on which limb is larger, and herein we will assume $\Delta R$ is positive, so that the example planet's morning limb is larger. While this construction simplifies the geometry, we note it is not a strictly realistic model, as it creates a discontinuity in the atmosphere at the planet's poles that grows with $\Delta R$. However, this simplified construction is likely adequate given current data quality, which we discuss later.

An observer seeking to study this planet's limb asymmetry must measure both $R_{\mathrm{evening}}$ and $R_{\mathrm{morning}}$\footnote{In practice, it is really the corresponding planet--star radius ratios $R_{\mathrm{evening}}/R_\star$ and $R_{\mathrm{morning}} / R_\star$ that are measured from a transit light curve.} as a function of wavelength, thus constructing separate transmission spectra for each limb. We assume that previous observers of this planet were not aware of -- or not capable of -- measuring limb asymmetry on the planet, and fit their data assuming the planet had uniform limbs. The planetary radius that these analyses will have inferred is the radius of a uniform circle with the same occulting area -- the same transit depth -- as the asymmetric-limb disk, which we call the ``effective radius" $R_{\mathrm{eff}}$, calculated as
\begin{equation}
    R_{\mathrm{eff}} = \sqrt{ \frac{1}{2} \left( R_{\mathrm{evening}}^2 + R_{\mathrm{morning}}^2 \right) }. \label{eqn:equivuniformradius}
\end{equation}
Figure~\ref{fig:exaggerated_intro_schematic} shows a schematic of the disk of this asymmetric-limb planet, shown in blue, compared to its uniform-limb equivalent, shown in red, at different points during transit. For reasons described further below, these previous analyses will also have inferred a time of conjunction, $t_{c, apparent}$, which is offset from the planet's true time of conjunction $t_{c, true}$ by some amount $\Delta t_c$ due to the planet's limb asymmetry.  In this work, we seek to understand the relationship between $\Delta R$ and $\Delta t_c$ and how it depends on the particular planet's properties.

To best intuit the degeneracy between $\Delta R$ and $\Delta t_c$, it is useful to look at the light curve of the asymmetric-limb planet compared to its uniform-limb counterpart. Figure~\ref{fig:exaggerated_intro_schematic} shows an exaggerated example of each at the bottom, which are aligned in time with the sequence of disk schematics at the top. In this example, both planets have the exact same true time of conjunction and orbital speed, and are moving from left to right across the stellar disk. The asymmetric-limb planet will begin to transit before the uniform-limb planet because its leading limb, which is $R_{\mathrm{morning}}$, is larger than $R_{\mathrm{eff}}$. In other words, because it has a larger limb, the asymmetric-limb disk makes first contact (T$_1$) with the stellar disk before the uniform-limb disk does. Similarly, since the asymmetric-limb disks' trailing edge, $R_{\mathrm{evening}}$, is smaller than $R_{\mathrm{eff}}$, it is still ``ahead" of the trailing edge of the uniform-limb disk, so it will also contact the stellar disk first during the end of ingress (T$_2$). The exact same sequence of events occurs during transit egress for the third (T$_3$) and fourth (T$_4$) contact points. Also note that the shape of the ingress and egress in each light curve is slightly different due to the different shape of each disk, but they both reach the same depth during full transit since both disks occult the same total area. 

Even though the two planets in Figure~\ref{fig:exaggerated_intro_schematic} have the same true time of conjunction, they have different contact times (T$_1$ - T$_4$) due to the difference in their limb radii, and this results in an apparent shift in the transit light curve along the time axis. That is, the blue light curve (of the asymmetric-limb planet) in Figure~\ref{fig:exaggerated_intro_schematic} looks like the red light curve (of the uniform-limb planet) but shifted slightly earlier in time. Using our notation, this means the uniform-limb planet's apparent time of conjunction $t_{c, apparent}$ is shifted slightly earlier by an amount $\Delta t_c$. 

This apparent shift in time is the source of the degeneracy between limb asymmetry and transit timing: the changes in the transit contact times due to limb asymmetry can be mimicked by varying the transit's time of conjunction and keeping the limb radii equal. The consequence for an observer is that while the blue light curve is the underlying astrophysical reality, literature analyses that do not account for potential limb asymmetry will infer the planet's time of conjunction to be $t_{c, apparent}$ rather than $t_{c, true}$. This presents a problem, because even if an observer obtains an ultra-high-precision light curve, if they fix the time of conjunction to $t_{c, apparent}$ in their model they will not be able to detect any limb asymmetry. Its effect has been canceled out by the systematic error ($\Delta t_c$) in the assumed transit time.

In the next section, we will derive a description for this degeneracy by solving for the relation between $\Delta R$ and $\Delta t_c$. We take advantage of the fact that, for a circular orbit in the uniform-limb case, the first and last contact times are equidistant from the time of conjunction. As a result, we can transform translations in $t_c$ to translations in T$_1$ and T$_4$ which can be geometrically related to the shape and size of the planetary disk. In essence, we are solving the problem of what $\Delta t_c$ is necessary to make the T$_1$ and T$_4$ of the uniform-limb light curve be the same as those of the asymmetric-limb light curve.

    \subsection{Times of First and Last Contact} \label{subsec:analytical_contact}

We assume that each planet rotates about a single axis that is oriented parallel to its orbital axis and lies in the plane of the page. The following derivation will hold reasonably well even if there is small rotation--orbital axis misalignment, though such misalignment is unlikely for short-period exoplanets, due to tidal interactions with the host star \citep[e.g.][]{hut1981_tidalevolution, correia2010_tidalevolution}. As shown in Figure~\ref{fig:exaggerated_intro_schematic}, we construct each planetary disk as two semicircles whose straight edges are joined along the planet's rotation axis. Here, we explicitly define the ``center" of each planet as the point where the rotation axis meets the two lines which bisect the arcs of each semicircle, as indicated in Figure~\ref{fig:exaggerated_intro_schematic}. Under this construction, these bisectors are parallel to the planets' transit trajectories. To aid our derivation, we will work in a Cartesian coordinate system as is also shown in Figure~\ref{fig:exaggerated_intro_schematic}. The x-axis is aligned along the planet's transit chord, so that the center of the planet is always moving directly along the x-axis. 

The point of conjunction is the position of the center of each planet at its time of conjunction, defined as when the center of the planet is closest to the center of the stellar disk. We define this point of conjunction to be the origin of our coordinate system, so that the y-axis passes through this point and is orthogonal to both planets' transit trajectories. Note that, in Figure~\ref{fig:exaggerated_intro_schematic}, we assume both planets transit directly along the stellar equator, but this need not be true in general. Our coordinate system is invariant to general changes in the planets' orbital orientations, as the x-axis can be translated vertically along the stellar disk and the y-axis rotated such that the x-axis is always horizontal. 

We assume the planets are of identical mass, which is negligible compared to the mass of the star, and are on identical circular orbits with semi-major axis $a$ and period $P$. The tangential speeds of each planet in their orbit are thus equal and given by
\begin{equation}
    v_p = \frac{2 \pi a}{P}. \label{eqn:orbit_velocity}
\end{equation}
Since these orbits are circular, $v_x \leq v_p$ as some motion is directed into or out of the page except at the point of conjunction. For simplicity, however, we will make the assumption that $v_x = v_p$ as this correction is generally small enough that it can be neglected. 

\begin{figure}
    \centering
    \includegraphics[width=\columnwidth]{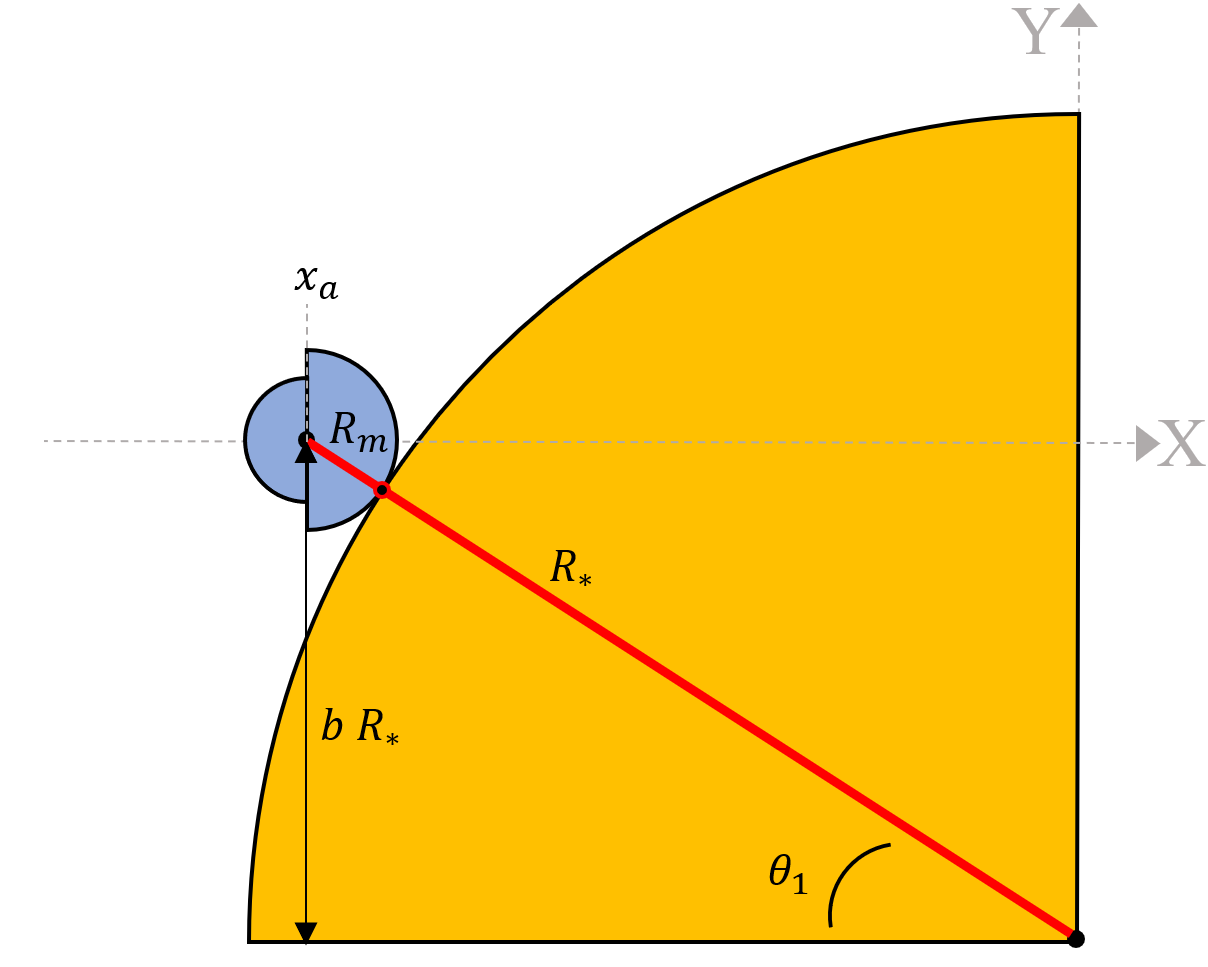}
    \caption{Projected geometry of a planet with asymmetric limbs at transit first contact. A line can be drawn from the center of the star to the center of the planet that passes through the point of first contact. This line defines the hypotenuse of a triangle that we use to derive how the time of first contact changes as the leading limb radius changes in Equation~\ref{eqn:dT1}.}
    \label{fig:firstcontacttriangle}
\end{figure}

When the asymmetric-limb planet makes first contact, assuming it has a larger leading limb, the uniform-limb planet will still need to travel some extra projected distance in the $x$-direction before it too makes first contact. Let us hold the asymmetric-limb planet fixed at first contact, and let the uniform-limb planet move forward in time until it too makes first contact. Then, we solve for the difference in these times via this extra projected distance $\Delta x$ that the uniform-limb planet travels. When the planets are transiting along the stellar equator, as is shown in Figure~\ref{fig:exaggerated_intro_schematic}, this is simply $\Delta x = R_{\mathrm{morning}} - R_{\mathrm{eff}} \approx \Delta R / 2$. For nonzero impact parameters, estimating $\Delta x$ is more involved. Figure~\ref{fig:firstcontacttriangle} shows the projected geometry at the point of first contact for an asymmetric-limb planet with nonzero impact parameter. We can draw a triangle with a hypotenuse from the center of the stellar disk, through the point of first contact, to the center of the planetary disk. The length of this hypotenuse is $R_\star + R_{\mathrm{morning}}$ and it makes an angle $\theta_1$ with respect to the stellar equator. The projected vertical distance between the center of the planet and the stellar equator is $b R_\star$, where $b$ is the impact parameter. Let us define the position of the center of the asymmetric-limb planet when its leading limb makes first contact as $x_a$, and likewise for the uniform-limb planet as $x_u$. The extra distance traveled by the uniform-limb planet is
\begin{dmath}
    \Delta x = x_a - x_u \\
    = \left(R_{\mathrm{morning}} + R_\star \right) \cos \left( \theta_{1, asym.} \right) \\- \left( R_{\mathrm{eff}} + R_\star \right) \cos \left( \theta_{1, unif.} \right). \label{eqn:dx_firstcontact_1}
\end{dmath}

For the triangle constructed in Figure~\ref{fig:firstcontacttriangle}, we can write 
\begin{eqnarray}
   \sin \left( \theta_{1,i} \right) = \frac{ b R_\star }{R_\star + R_{p,i} }. \label{eqn:ctri_sine}
\end{eqnarray}
Here, $R_{p,i}$ can be $R_{\mathrm{morning}}$ or $R_{\mathrm{eff}}$. We can convert Equation~\ref{eqn:ctri_sine} to the cosine of the angle
\begin{eqnarray}
   \cos \left( \theta_{1,i} \right) = \sqrt{ 1 - \left( \frac{ b R_\star }{R_\star + R_{p,i}} \right)^2 }, \label{eqn:ctri_cosine}
\end{eqnarray}
and thus
\begin{eqnarray}
   x_i &=& \left( R_\star + R_{p,i} \right) \sqrt{ 1 - \left( \frac{ b R_\star }{R_\star + R_{p,i}} \right)^2 } \\
       &=& \sqrt{ \left( R_\star + R_{p,i} \right)^2 - \left(b R_\star \right)^2 }. \label{eqn:xi_simplified}
\end{eqnarray}
We can then rewrite Equation~\ref{eqn:dx_firstcontact_1} as
\begin{dmath}
   \Delta x = \sqrt{ \left( R_\star + R_{\mathrm{morning}} \right)^2 - \left(b R_\star \right)^2 } \\- \sqrt{ \left( R_\star + R_{\mathrm{eff}} \right)^2 - \left(b R_\star \right)^2 }.
\end{dmath}
In the limiting case of $b = 0$, this simplifies back to $\Delta x  = R_{\mathrm{morning}} - R_{\mathrm{eff}}$. Then, since $\Delta T_i = \Delta x / v_p$, the difference in the times of first contact is
\begin{dmath}
   \Delta T_1 = \frac{P}{2 \pi a} \left(  \sqrt{ \left( R_\star + R_{\mathrm{morning}} \right)^2 - \left(b R_\star \right)^2 } - \sqrt{ \left( R_\star + R_{\mathrm{eff}} \right)^2 - \left(b R_\star \right)^2 } \right). \label{eqn:dT1}
\end{dmath}

The same derivation can be done for the time of fourth contact by swapping the morning limb radius for the evening limb radius. In our construction, the uniform-limb planet will also make fourth contact after the asymmetric-limb planet, now because the uniform-limb radius is larger than the evening-limb radius. Changing out terms, we then have
\begin{dmath}
   \Delta T_4 = \frac{P}{2 \pi a} \left(  \sqrt{ \left( R_\star + R_{\mathrm{eff}} \right)^2 - \left(b R_\star \right)^2 } - \sqrt{ \left( R_\star + R_{\mathrm{evening}} \right)^2 - \left(b R_\star \right)^2 } \right). \label{eqn:dT4}
\end{dmath}
By construction, $\Delta T_1$ and $\Delta T_4$ will always have the same sign. This is because the equivalent-depth uniform radius will always be between the two limb radii, so T$_1$ and T$_4$ will always be earlier or later together. If the asymmetric-limb planet's morning limb is larger than its evening limb, as we have been using in our constructed example, then $\Delta T_1$ and $\Delta T_4$ will be negative. If the evening limb is larger, they will be positive. Then, because these contact times are equidistant about the time of conjunction for the uniform-limb planet: 
\begin{dmath}
\Delta t_c = \frac{1}{2} \left( \Delta T_1 + \Delta T_4 \right). \label{eqn:dt_analytic}
\end{dmath}

\section{Numerical Method} \label{sec:numerical}
    \subsection{Setup} \label{subsec:numerical_setup}

We simulated transit observations of the exoplanets WASP-107b, WASP-39b, WASP-96b, and WASP-80b to compare our analytic formulae in Section~\ref{subsec:analytical_contact} to numeric models. We chose these specific planets because they are all JWST Early Release Science and/or Cycle 1 observation targets, and will likely be immediate targets for investigation of limb asymmetry. Further, these planets represent a wide range of bulk atmospheric scale heights, which we use as a measure of the magnitude of limb asymmetry. For these simulations, we collected literature values for several planetary and orbital parameters, which are listed in Table~\ref{tab:planet_properties}. We computed the bulk atmospheric scale height $H$ of each planet based on the published mass, radius, and equilibrium temperature, and an assumed atmospheric mean molecular weight of 2.3 atomic mass units for each planet. We then followed our previous construction by setting the planet's evening limb radius $R_{p, evening}$ to the published radius, then compute the morning limb radius as
\begin{dmath}
    R_{p, morning} = R_{p, evening} + N \times H.
\end{dmath}
Here, $N$ is any positive number and $N \times H$ takes the place of $\Delta R$ so that the radii of the evening and morning limbs differ by some multiple of the planet's bulk scale height. 

We then simulated one transit observation of each planet. We first generated a model light curve of the asymmetric-limb planet using the transit modeling package \texttt{catwoman} \citep{Espinoza2021_catwoman2, catwoman1} with literature values for the orbital parameters, the above prescription for the limb radii assuming some value for $N$, uniform limb darkening, and assuming the axis separating the limbs is aligned to the orbital axis. We use this model light curve as a reference for generating synthetic observed data, computing the relative flux value of each synthetic data point as 
as
\begin{equation}
    y_{data, i} = y_{model, i} + \mathcal{G} \left(\mu = 0, \sigma \right).
\end{equation}
Here, we emulate photon noise in the light curve by adding pure white noise to these data via the function $\mathcal{G} \left(\mu = 0, \sigma \right)$, which represents draws from a Gaussian distribution with a mean of zero and a standard deviation of $\sigma$. We assume this photon noise is the only source of scatter in the light curve. 

We first simulated a ``perfect" transit observation of the exoplanet WASP-39b in order to directly compare to our analytical formulae. We generated a noiseless light curve with data points at a cadence of 0.5~s. For this ``perfect" observation, we finely sampled $N$ in 20 evenly spaced steps between 0 and 5. 

\begin{table*}
    \begin{centering}
    \caption{Planetary and Stellar Properties Used in Our Transit Observation Simulations.}
    \begin{tabular}{|c|c|c|c|c|c|}
    Parameter (units) &  WASP-80b & WASP-96b & WASP-39b & WASP-107b \\ \hline 
         P (days)                     &  3.06785271(19)\tablenotemark{a}    & 3.4252565(8)\tablenotemark{a}     & 4.0552941(34)\tablenotemark{f}     & 5.721488(3)\tablenotemark{a} \\
         a / R$_s$                    & 12.63 $\pm$ 0.13\tablenotemark{b}   & 9.03 $\pm$ 0.30\tablenotemark{d}  & 11.55 $\pm$ 0.13\tablenotemark{g}  & 18.2 $\pm$ 0.1\tablenotemark{j} \\
         $i$ (degrees)                & 89.02 $\pm$ 0.11\tablenotemark{b}   & 85.6 $\pm$ 0.2\tablenotemark{c}   &  87.32 $\pm$ 0.17\tablenotemark{f} & 89.7 $\pm$ 0.2\tablenotemark{j} \\
         R$_p$ / R$_\star$            &  0.17137(39)\tablenotemark{b}       & 0.1186(17)\tablenotemark{d}       & 0.1457(15)\tablenotemark{h}        & 0.14434(18)\tablenotemark{k} \\
         R$_\star$ (R$_\odot$)        &  0.605 $\pm$ 0.048\tablenotemark{*} & 1.15 $\pm$ 0.03\tablenotemark{*}  & 0.92 $\pm$ 0.03\tablenotemark{*}   & 0.73 $\pm$ 0.15\tablenotemark{*} \\
         M$_p$ (M$_{\text{Jupiter}}$) &  0.538 $\pm$ 0.035\tablenotemark{b} & 0.49 $\pm$ 0.04\tablenotemark{e}  & 0.281 $\pm$ 0.032\tablenotemark{f} & 0.096 $\pm$ 0.005\tablenotemark{l} \\
         T$_{eq}$ (K)                 &  825 $\pm$ 19\tablenotemark{b}      & 1285 $\pm$ 40\tablenotemark{c}    & 1166 $\pm$ 14\tablenotemark{f}     & 770 $\pm$ 60\tablenotemark{j} \\
         T$_{14}$ (hr)             &  2.131 $\pm$ 0.003\tablenotemark{b} & 2.4264 $\pm$ 0.0264\tablenotemark{c} & 2.8032 $\pm$ 0.0192\tablenotemark{i}    & 2.753 $\pm$ 0.007\tablenotemark{j} \\
         H (km)                       &  226 $\pm$ 44                       & 669 $\pm$ 97                      & 1042 $\pm$ 171                     & 1221 $\pm$ 174     \\
         Cadence (s)                  &  14.8                               & 72.0                              & 79.5                               & 20.2          \\
         $\sigma$ (ppm)               &  200                                & 150                               & 200                                & 200         
    \end{tabular}
    \label{tab:planet_properties}
    \end{centering}
    \tablecomments{
    Uncertainties with many digits are given in parenthetical form, where the values of $N$ digits in each parenthesis replace the last $N$ significant digits of the measured value. For example, 0.03(1) means 0.03 $\pm$ 0.01. Values without associated references were either assumed or calculated based on other values. 
    For each planet, not all parameters were readily available from a single reference. Strict self-consistency of these values does not impact our results, so in these cases we combine measurements from multiple references as convenient.
    Parameter definitions: $P$ = orbital period, $a$ = semi-major axis, $R_\star$ = stellar radius, $i$ = orbital inclination, $R_p$ = planetary radius, $M_p$ = planetary mass, $T_{eq}$ = planetary equilibrium temperature, $T_{14}$ = transit duration, $H$ = bulk atmospheric scale height, Cadence = point-to-point time-step in synthetic data, $\sigma$ = synthetic light-curve scatter.}
    \tablerefs{a-\cite{kokori2022}, b-\cite{triaud2015}, c-\cite{hellier2014},
    d-\cite{patel2022}, e-\cite{bonomo2017}, f-\cite{mancini2018}, g-\cite{fischer2016},  h-\cite{maciejewski2016}, i-\cite{faedi2011}, j-\cite{anderson2017}, 
    k-\cite{dai2017}, l-\cite{piaulet2021}, *-\cite{gaiadr2}
    }
\end{table*}

We then simulated more realistic transit observations of each planet using the optimal integration cadences computed using \texttt{PandExo} \citep{pandexo} for a JWST/NIRCam F322W2 observation, and light-curve scatters of 200~ppm. For these simulations, we used a coarser sampling for $N$ of 10 steps between 0 and 5. The cadences are given in Table~\ref{tab:planet_properties} and ranged from $\sim$15 - 80 s, consistent with those used by real observations of these targets \citep{bell2023_wasp80b, ahrer2023_wasp39b, murphy:inreview}. This light-curve scatter is slightly optimistic for JWST/NIRCam F322W2, as real observations of these targets have achieved $\sim$230-500 ppm scatter \citep{bell2023_wasp80b, carter:inprep, murphy:inreview}, but it has been achieved by other JWST instruments like NIRISS SOSS \citep{radica2023_wasp96b}. For WASP-96b, its nominal cadence of 144~s and 200~ppm light-curve scatter led to large noise in our simulation results. We therefore used a 2$\times$ faster cadence of 72~s and slightly smaller scatter of 150~ppm just for WASP-96b, which did not change the underlying result but made it easier to visually present and compare to our analytical results. In all cases, we assumed having an equal time out-of-transit baseline to the published transit duration.

After generating the simulated observed data, we fit these data using a uniform-limb transit model generated by the transit modeling package \texttt{batman} \citep{batman}, also with uniform limb darkening. We left all parameters fixed to their true values except for the time of conjunction, which was allowed to vary freely. We set the planetary radius to the value that would have the same planetary disk occulting area as the corresponding asymmetric-limb disk, using Equation~\ref{eqn:equivuniformradius}. In the ``perfect" noiseless case, we simply fit the data through a least-squares regression. Then in the other cases, we fit the data using the Markov Chain Monte Carlo method using \texttt{emcee} \citep{emcee}, sampling just the time of conjunction for 5500 steps which was more than sufficient for each run's sampling to converge. 

We chose to use uniform limb darkening when generating the reference light curves, meaning the stellar disks are completely uniform in brightness. This choice was partially motivated by computational efficiency, as the overhead initializations of \texttt{catwoman} and \texttt{batman} are significantly slower when limb darkening must be computed. It was also motivated by our desire to treat each planet uniformly, as different host stars exhibit different limb darkening that may yield different results depending on the assumed cadence or noise of the light curve. We did test various limb darkening laws, including quadratic and logarithmic, and various sets of limb darkening coefficient values. We found that the precise choice of limb darkening did not change our result so long as it was well known, because we set the limb darkening when generating the reference light curve and kept it fixed for all later steps, thus assuming perfect knowledge of limb darkening during the light-curve fitting. Uncertainty in the limb darkening coefficients is of concern to real observations, but is beyond the scope of our analysis and can be mitigated using stellar models in practice.

    \subsection{Connection to Literature Measurements} \label{subsec:numerical_literatureconnection}

Our numerical method described in Section~\ref{subsec:numerical_setup} resembles the method by which most, if not all, literature measurements of exoplanet transit times are determined. In reality, one obtains a transit observation of an exoplanet that may or may not have asymmetric limbs. The data are then typically fit using a uniform-limb model for the planet's disk, using \texttt{batman} or similar light-curve models, typically using a Bayesian parameter sampling technique, such as \texttt{emcee}. Therefore, using this numerical method we are able to directly compare our analytical predictions to what one may derive from a real observation.
    
\section{Comparing Our Analytical and Numerical Results} \label{subsec:numerical_verification}

\begin{figure*}
    \centering
    \includegraphics[width=\textwidth]{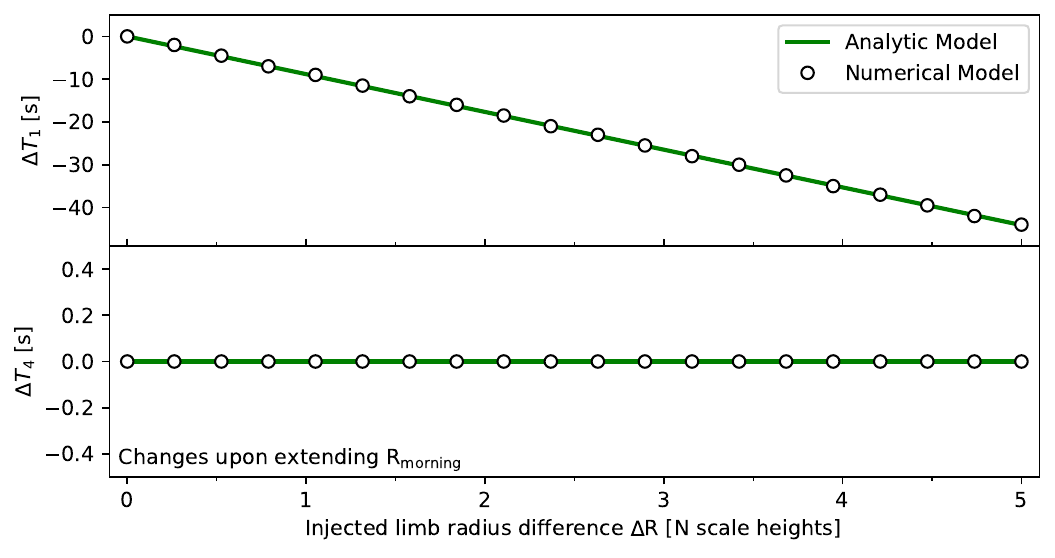}
    \caption{Changes in the time of first ($\Delta T_1$) and last ($\Delta T_4$) contact due to extending the radius of our example planet's morning limb relative to the evening limb. We defined this radius extension $\Delta R$ in multiples of the planet's atmospheric scale height, as given on the x-axis. Note that we change only the morning limb radius -- i.e., the leading limb -- so only T$_1$ is affected. T$_4$ should not be affected. 
    For this example, we used literature parameters for the planet WASP-39b and generated light curves using the transit model \texttt{catwoman} for each $\Delta R$. Then, we recorded how T$_1$ and T$_4$ of each light curve changed relative to the case of the originally uniform limb sizes ($\Delta R$ = 0), shown as the black points. We compare these results to our analytical predictions for $\Delta T_1$ and $\Delta T_4$, given in Equations~\ref{eqn:dT1} and \ref{eqn:dT4}, which we find perfectly match the numerical results.}
    \label{fig:perfectLC_selfT1T4changes}
\end{figure*}

First, we recorded the reference light curves generated for the ``perfect" observation, determined the first (T$_1$) and last (T$_4$) contact times of each light curve, and calculated the change in both relative to the $N$ = 0 (i.e. uniform-limb) case as $N$ increased. Note the slight difference that this is comparing an increasing $R_{\mathrm{morning}}$ to the original $R_{\mathrm{evening}}$, and not comparing to a covarying $R_{\mathrm{eff}}$, so only T$_1$ will change. We compared these changes to what Equations~\ref{eqn:dT1} and \ref{eqn:dT4} predict, and found they perfectly agree, as shown in Figure~\ref{fig:perfectLC_selfT1T4changes}. In terms of magnitude, we found that T$_1$ occurs over 40~s earlier when the morning limb radius is increased by five scale heights. As a further test, we found the same behavior for T$_4$ upon adjusting only the evening limb radius.

\begin{figure*}
    \centering
    \includegraphics[width=\textwidth]{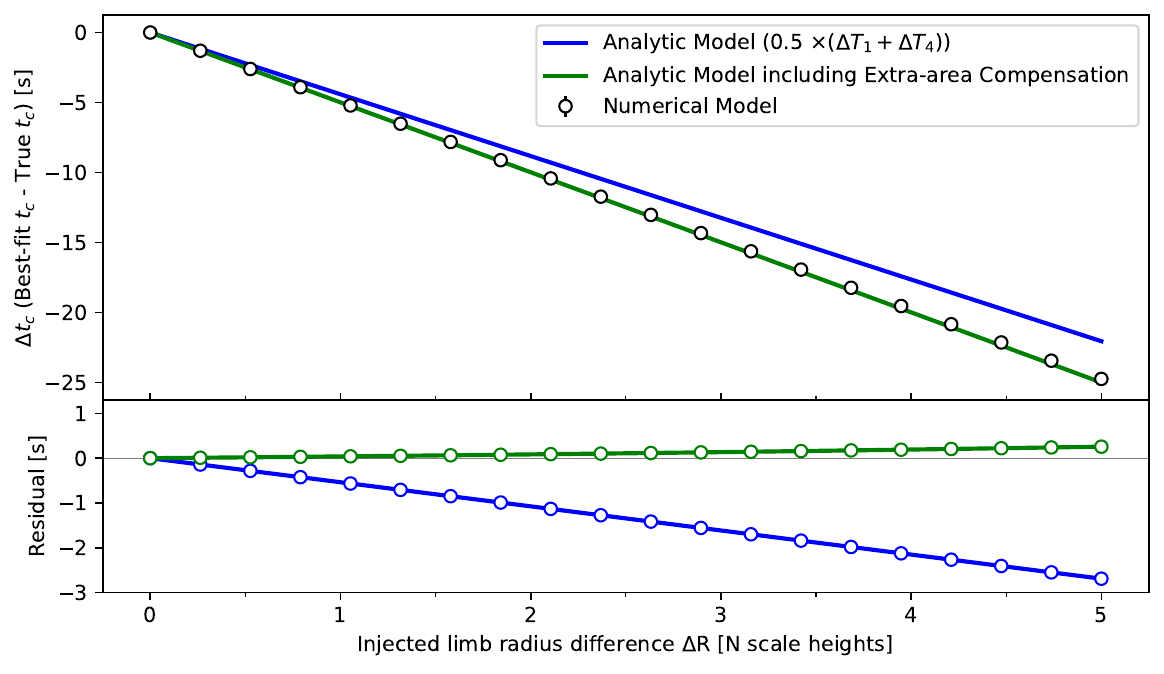}
    \caption{Measured times of conjunction compared to the true times of conjunction, as a function of the injected difference in radius between the planet's evening and morning limbs. The synthetic data used in these fits use an asymmetric-limb transit model. The circular points show fitting results from our numerical model, which fits using a uniform-limb transit model. The blue line shows our analytical model's prediction for this bias in $t_c$ (Equation~\ref{eqn:dt_analytic}). The discrepancy between the numerical model and the analytical model is due to the ``extra-area compensation effect," described in Section~\ref{subsec:numerical_extraareaproblem}. The green line shows the analytic model when including this compensation effect (Equation~\ref{eqn:compensation_general_tau_lineend}).
    }
    \label{fig:perfectLC_timingbias}
\end{figure*}

Next, we computed the difference between the best-fit time of conjunction of the uniform-limb model for the ``perfect" observation's simulated data to the true value, $\Delta t_c$. This is shown as the circular points in Figure~\ref{fig:perfectLC_timingbias}. We compared this numerical time offset to our analytical model's prediction for $\Delta t_c$ (Equation~\ref{eqn:dt_analytic}), shown as the blue line. The numerical model found a larger $\Delta t_c$ than the analytic model by a factor of about 10\% at all $N$, reaching nearly a 3 second discrepancy at our maximal case of limb asymmetry of $N$ = 5. This discrepancy consistently appeared, with approximately the same relative magnitude, in all of our numerical vs. analytical comparisons regardless of our specific choice of planetary parameters, limb darkening coefficients, and noise parameters. We discuss the underlying reason for this in the following section.

    \subsection{Extra-area Compensation When fitting transits} \label{subsec:numerical_extraareaproblem}

The discrepancy between the time offsets estimated by our numeric model and our analytic model (Figure~\ref{fig:perfectLC_timingbias}) stems from an inherent difference in what each aims to measure. Our analytic model was derived to specifically calculate the $\Delta t_c$ required to make the T$_1$ and T$_4$ of the asymmetric-limb and uniform-limb planets match. In the numeric model, we are instead calculating the ``best-fit" $t_c$ of the uniform-limb model when it is applied to data generated from an asymmetric-limb model. The ``best-fit" solution is determined not strictly by making the T$_1$ and T$_4$ values match, but by the specific value of $t_c$ that optimizes the goodness-of-fit metric used by the fitting routine. In our case, this metric was either the sum of squared residuals or a Gaussian likelihood function. Fits to real, noisy data often use the latter. 

\begin{figure*}
    \centering
    \includegraphics[width=0.9\textwidth]{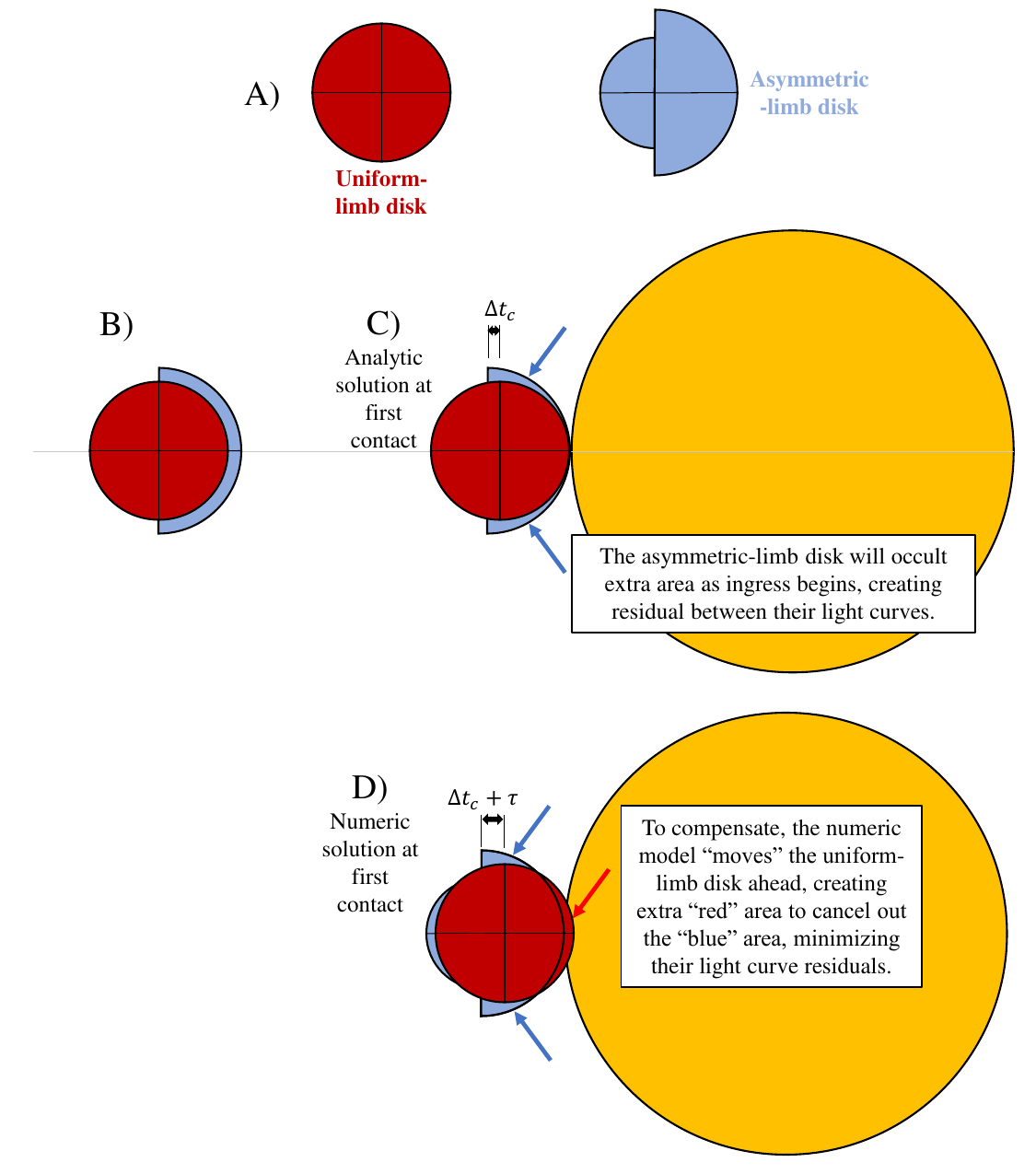}
    \caption{
    Schematic of the extra-area compensation problem. We generate a synthetic transit observation using the asymmetric-limb disk, shown in blue (Panel~A). We fit these data using the uniform-limb disk shown in red. These two model planets are initially set to have the same time of conjunction $t_c$ (Panel~B). Due to limb asymmetry imprinted into the data, this fit will return a biased $t_c$ when using this uniform-limb model. Panel~(C) shows our analytic solution for this bias ($\Delta t_c$), which is the extra time the uniform-limb disk would travel until the two disks' first and last contact times are equal. This value of $\Delta t_c$ is deemed a poor fit by the numeric model, however, as the ``extra" occulting area of the asymmetric-limb disk in this arrangement leads to a large fitting residual. Rather, the numeric model prefers the solution shown in Panel~(D) where the uniform-limb disk is translated even further ahead by time $\tau$. In this arrangement, there is now ``extra" occulting area for the uniform-limb disk, indicated by the red arrow, which cancels out the asymmetric-limb disk's extra area, leading to a minimized residual and thus a ``better" fit. }
    \label{fig:extra_area_schematic}
\end{figure*}

We find that the discrepancy in time offsets results from the numeric model preferring larger values of $\Delta t_c$ that serve to minimize the residuals in the fits to our simulated data. This is a purely numerical effect, which we dub the ``extra-area compensation effect". We illustrate how the effect arises in Figure~\ref{fig:extra_area_schematic}. In short, our analytic model solves for the $\Delta t_c$ that makes T$_1$ and T$_4$ match between the uniform-limb and asymmetric-limb planets, as shown in Panel~C. However, due to the difference in size and shape of the disks, the asymmetric-limb planet still has ``extra" occulting area, indicated by the blue arrows. This leads the asymmetric-limb planet to have a steeper ingress and egress compared to the uniform-limb planet, causing a large residual between their transit light curves despite having the same contact times. The numeric model compensates for this with an extra offset $\tau$ to the time of conjunction, which has the effect of eliminating this ``extra" area, shown in Panel~D. In other words, the numerical model is drawn toward the flux-weighted mid-transit time.

The planetary disks represented by our model light curves are shown for reference in panel~A, with the uniform-limb planet's disk in red and the asymmetric-limb planet's disk in blue. Both disks have the same total area. Panels~B, C, and D illustrate the planet approaching transit, with the star represented as the uniform yellow-orange disk. We start by treating the planets as having the same $t_c$, shown in panel~B. Due to the limb asymmetry, fitting the uniform-limb model to an observation of the asymmetric-limb planet would lead to an offset $\Delta t_c$ in the inferred time of conjunction. This arises because the uniform disk's time of first contact lags behind that of the asymmetric disk's by time $\Delta t_c$. Our analytic model solves for this value of $\Delta t_c$ (Panel~C). In our analytic model's solution, however, the asymmetric disk has two regions of ``extra" occulting area -- due to its larger leading limb -- which are indicated by the blue arrows in Panel~C. Note that there is an equivalent ``missing" area on the other side of the disk that affects egress. During the first half of ingress, this ``extra" area leads to the asymmetric disk having a steeper ingress compared to the uniform disk, causing a large residual between the two light curves despite the two disks making first contact at the exact same time. Due to this residual, our numeric model disfavors this solution. The numeric model compensates for this ``extra" area by offsetting the $t_c$ of the uniform disk by an extra amount $\tau$. As shown in Panel~D, this has the effect of moving the edge of the uniform disk ahead of the asymmetric disk, so that the uniform disk now has its own ``extra" area indicated by the red arrow. The preferred $\tau$ is that which makes the red and blue extra areas equivalent. The asymmetric disk's steeper ingress is compensated by the uniform disk beginning ingress earlier, which has the net effect of minimizing the residual between their light curves. As a result, the ``best-fit" solution from the numerical model is the situation shown in Panel~D, where the uniform-limb model's time of conjunction is offset by $\Delta t_c + \tau$.

\begin{figure}
    \centering
    \includegraphics[width=\columnwidth]{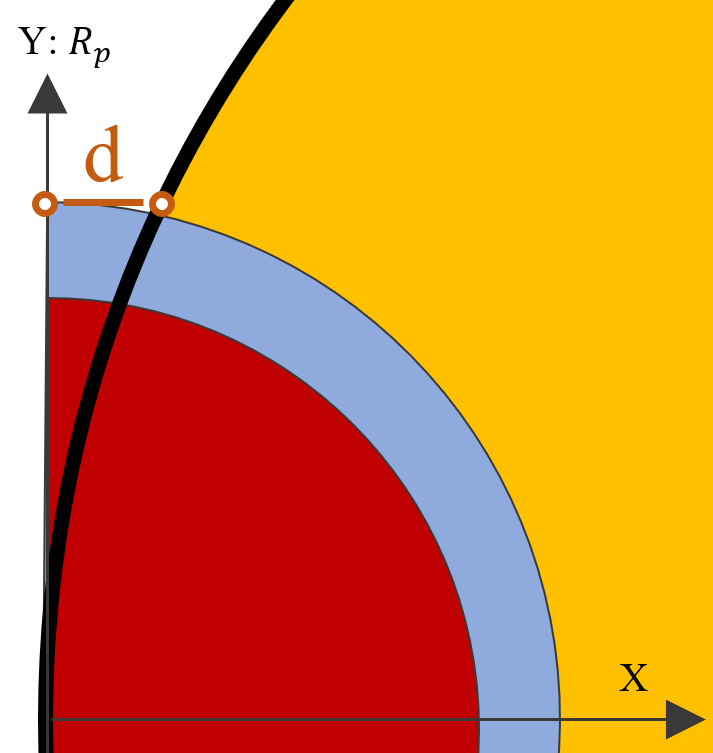}
    \caption{Schematic showing the extra distance necessary for a disk hemisphere to travel between the point when the whole-disk center is contacting the disk and the point when the entire hemisphere will be internally tangent. The blue and red semicircles represent the hemispheres of our uniform-limb and asymmetric-limb planets shown in Figures~\ref{fig:exaggerated_intro_schematic} and \ref{fig:compensation_derivation_schematic}, and the orange circle is the star. This extra distance is equal to the $x$-value of the stellar disk at a given $y$, where the $y$-coordinate is described by the radius of the disk hemisphere. We use this as the basis for our derivation of the extra-area compensation given by Equation~\ref{eqn:compensation_general_tau_lineend}. 
    }
    \label{fig:compensation_derivation_schematic}
\end{figure}

We can derive an analytical estimate for this compensation $\tau$ by comparing the path lengths traveled by the leading halves of each disk during ingress. Consider the alignment shown in Panel~B of Figure~\ref{fig:extra_area_schematic}, where the center of each disk is aligned. We position each disk halfway through ingress where the center of the disk makes contact with the edge of the stellar disk, as illustrated in Figure~\ref{fig:compensation_derivation_schematic}. We only show the leading half of each disk for simpler visualization. At this point, each half-disk must still travel some additional distance $d$ along $x$ before that half-disk is fully internally tangent (i.e., fully inside the stellar disk). For example, this additional distance for the asymmetric disk is given by the orange line at the top of Figure~\ref{fig:compensation_derivation_schematic}. The area of the planetary disk still remaining outside the stellar disk is analogous to the ``extra" areas discussed above. In the coordinate system shown in Figure~\ref{fig:compensation_derivation_schematic}, the distance between the top of each planetary disk and the edge of the stellar disk is equal to 
\begin{dmath}
    d = R_\star - \sqrt{ R_\star^2 - y^2}. \label{eqn:y_defining_stellardisk}
\end{dmath}
Note that $y$ also represents a length in this equation. When each disk has traversed approximately $d$/2, an area equal to what still remains outside the stellar disk will have entered within the stellar disk. This is analogous to the ``extra" area compensation done by the numerical model, as shown in Panel~D of Figure~\ref{fig:extra_area_schematic}. Therefore, we can estimate $\tau$ from the difference in $d$/2 between the uniform-limb and asymmetric-limb disks, $\Delta d$/2. Based on Equation~\ref{eqn:y_defining_stellardisk}, this depends on the difference of their radii and the (fixed) radius of the star: 
\begin{eqnarray}
    \frac{\Delta d}{2} &=& \frac{1}{2} \left[ d_1 - d_2 \right] \\
                       &=& \frac{1}{2} \left[ \left( R_\star - \sqrt{ R_\star^2 - y_1^2} \right) - \left( R_\star - \sqrt{ R_\star^2 - y_2^2} \right) \right]. \\
                       &=& \frac{1}{2} \left[ \sqrt{ R_\star^2 - y_2^2}  - \sqrt{ R_\star^2 - y_1^2}  \right].
\end{eqnarray}
Here, $y_1$ and $y_2$ represent the polar radii of the larger and smaller half-disks, respectively. In our case, these are the radii of the asymmetric disk's leading (i.e., morning) edge and the uniform disk, so $y_1 = R_{\mathrm{morning}}$ and $y_2$ = $R_{\mathrm{eff}}$. To convert this to a duration of time, we simply divide by the (fixed) orbital speed. The result is
\begin{eqnarray}
    \tau &=& \frac{1}{v_p} \frac{\Delta d}{2} \label{eqn:compensation_general_tau_line1} \\
    \tau &=& \frac{1}{2 v_p} \left[ \sqrt{ R_\star^2 - y_2^2}  - \sqrt{ R_\star^2 - y_1^2}  \right] \label{eqn:compensation_general_tau_lineend} \\
    \tau &=& \frac{1}{2 v_p} \left[ \sqrt{ R_\star^2 - R_{\mathrm{eff}}^2}  - \sqrt{ R_\star^2 - R_{\mathrm{morning}}^2}  \right]. \label{eqn:compensation_tau_Rpexample}
\end{eqnarray}

We computed the expected extra-area compensation for our ``perfect" observation of WASP-39b scenario using Equation~\ref{eqn:compensation_tau_Rpexample} and added it to our nominal analytical results shown in Figure~\ref{fig:perfectLC_timingbias}. The result is shown as the green line in Figure~\ref{fig:perfectLC_timingbias} which agrees well with our numerical model. There is a small residual between the two, which slightly increases with increasing $N$. As alluded to previously, this is likely the result of us assuming the asymmetric-limb planet's cross-section is composed of two connected semicircles with discontinuity in the polar regions, as this discontinuity becomes more extreme with increasing $N$, among other assumptions discussed in Section~\ref{subsec:discussion_modellimitations}. This residual is less than one quarter of a second for the most extreme case of limb asymmetry. Achieving such precision on $t_c$ is beyond the capability of current instruments, and limb asymmetry of $N>5$ may be physically unrealistic.

    \subsection{``Realistic" Case Results} \label{subsec:numerical_realisticresults}

\begin{figure*}
    \centering
    \includegraphics[width=\textwidth]{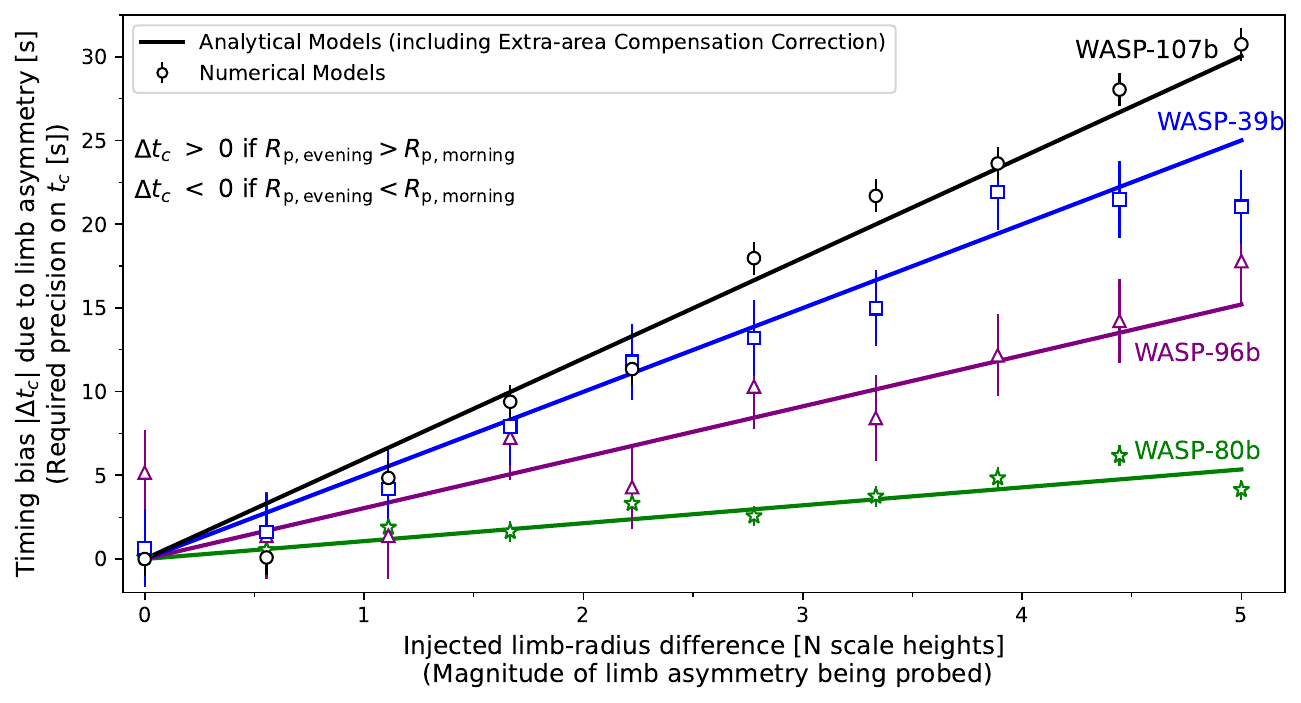}
    \caption{Differences between the measured times of conjunction compared to the true time of conjunction ($\Delta t_c$) as a function of the injected limb radius difference for simulated observations of representative exoplanets. In practice, these constitute the minimum a priori precision on $t_c$ necessary before limb asymmetry can be successfully measured from a transit observation. The solid lines show our analytical model's predictions for $\Delta t_c$ using Equation~\ref{eqn:dt_analytic}. The points show our numerical model, which mimics fitting real observations. We include the extra-area compensation effect in our analytic model using Equation~\ref{eqn:compensation_tau_Rpexample}. The key difference between the planets shown here is their atmospheric scale height. Higher scale height leads to larger $\Delta t_c$, meaning less a priori precision on $t_c$ is required to probe limb asymmetry. 
    }
    \label{fig:realisticLCs_tcbias}
\end{figure*}

Now we return to the results of our ``realistic" simulated observations of WASP-80b, WASP-96b, WASP-39b, and WASP-107b described in Section~\ref{subsec:numerical_setup}. We plot the best-fit timing biases $\Delta t_c$ as a function of $N$ for each simulation in Figure~\ref{fig:realisticLCs_tcbias}. We compare these to the corresponding analytic model predictions for each planet from Equation~\ref{eqn:dt_analytic} including the extra-area compensation from Equation~\ref{eqn:compensation_tau_Rpexample}, and find they agree extremely well. The residuals for each planet are consistent with being due to the noise and cadence of each simulated observation. The key meaning of these timing biases is that they represent the minimum precision on a planet's $t_c$ necessary to be able to probe limb asymmetry.

For each planet tested, the timing bias increases linearly from zero with increasing limb asymmetry. For modest ($N$ = 1-2) amounts of limb asymmetry, the timing bias ranges from $\sim$1 - 10~s, which is small but detectable -- especially with JWST observations. Note that, in Figure~\ref{fig:realisticLCs_tcbias}, we give the absolute values for $\Delta t_c$. As mentioned in Section~\ref{subsec:analytical_contact}, the sign of $\Delta t_c$ depends on which limb is larger -- $\Delta t_c >$ 0 when the evening limb is larger, and  $\Delta t_c <$ 0 when the morning limb is larger. We find that the slope of each planet's timing bias model differs and is correlated with the planet's bulk scale height, meaning planets with larger scale height have a larger timing bias for a given value of $N$.

\section{Discussion} \label{sec:discussion}
    \subsection{Implications for Investigating Limb Asymmetry}
    \label{subsec:discussion_limbasym}

Error in a planet's time of conjunction is a key systematic obstacle to investigating limb asymmetry via transmission spectroscopy. This raises the question of how accurately a planet's time of conjunction needs to be known in order to reliably probe limb asymmetry on the planet. Answering this question was the primary motivation for our work, and our models for the timing bias (Section~\ref{sec:analytical}, and Figure~\ref{fig:realisticLCs_tcbias}) provide aid for future analyses in understanding and overcoming this obstacle. 

The timing biases shown in Figure~\ref{fig:realisticLCs_tcbias} and calculated with Equations~\ref{eqn:dt_analytic} and \ref{eqn:compensation_general_tau_lineend} are the a priori measurement precision on $t_c$ necessary to measure limb asymmetry at a given scale height. In other words, if the uncertainty on the target's $t_c$ is larger than the timing bias for a given $N$, then probing limb asymmetry $\leq$$N$ is unreliable.

This relationship between the timing bias and the magnitude of limb asymmetry clearly depends on the particular properties of the target, as seen by the different slopes of each model in Figure~\ref{fig:realisticLCs_tcbias}. The values of these slopes are mostly determined by the planet's bulk atmospheric scale height, as this sets the scale of the difference in radius between limbs. To generalize our analytical model from Section~\ref{sec:analytical}, including the extra-area compensation effect from Section~\ref{subsec:numerical_extraareaproblem}, we fit the slope of each model shown in Figure~\ref{fig:realisticLCs_tcbias} as a linear function of the corresponding planet's scale height.  As a result, knowing only the target planet's atmospheric scale height $H$, the uncertainty on $t_c$ necessary to probe limb asymmetry of $N$ scale heights on a potential target can be quickly estimated as
\begin{eqnarray}
    \sigma_{t_c}~\mathrm{[s]} = \left[  0.00496 \left( \frac{H}{1~km} \right) - 0.13 \right] ~ N. \label{eqn:generaltimingbiaseqn}
\end{eqnarray}

We note that our analysis has considered a transit observation using only a single filter, such as a photometric or band-integrated spectroscopic observation. \cite{powell2019} showed the utility of instead using a multiwavelength approach so that the spectral variation of limb asymmetry, when combined and forced to share a single $t_c$, actually works against itself to eliminate timing biases. Our results are complementary to such an approach. This multiwavelength method will still lead to a best-fit $t_c$ with uncertainty, and our results demonstrate how well limb asymmetry may be explored with this remaining uncertainty. 

Similarly, an alternate approach would be to simply measure the relative changes in $t_c$ as a function of wavelength. These chromatic variations, especially if aligned with spectral features of expected molecules in the planet's atmosphere, would indicate the presence of limb asymmetry on the planet without needing an absolute $t_c$ for reference at all. While this approach indeed could easily identify the presence of a difference in atmospheric properties between limbs without concern of the degeneracy presented in this work, any deeper characterization of the limb-to-limb differences -- for example, fitting the transmission spectra of each limb across multiple instruments -- would require an absolute $t_c$ value.

    \subsection{Implications of the Extra-area Compensation Effect}\label{subsec:discussion_extraareaproblem}

In Section~\ref{subsec:numerical_extraareaproblem}, we described an extra offset in $t_c$ measured for an asymmetric-limb planet due purely to numerical effects during the fitting process. The magnitude of this offset depends on the difference between the morning and evening radii (Equation~\ref{eqn:compensation_tau_Rpexample}, recalling the definition of $R_{\mathrm{eff}}$ from Equation~\ref{eqn:equivuniformradius}). Therefore, per multiple of the scale height, this offset will also be larger for planets with larger scale heights, as there is more physical occulting area to compensate for. With this in mind, we can also generalize Equation~\ref{eqn:compensation_tau_Rpexample} by calculating $\tau$ as a function of $N$ for each of our example planets, which are linear functions, and fitting their slopes as a function of the scale height. We find that this extra-area compensation offset can therefore be approximated by
\begin{eqnarray}
    \tau~ \mathrm{[s]} = \left[ 0.00063~\left(\frac{H}{1~km}\right) - 0.05 \right] ~ N. \label{eqn:generalcompensationeqn}
\end{eqnarray}
Note that this effect is already included in the general formula given in Equation~\ref{eqn:generaltimingbiaseqn}. Therefore, subtracting Equation~\ref{eqn:generalcompensationeqn} from \ref{eqn:generaltimingbiaseqn} gives a general estimate for our analytical model from Section~\ref{sec:analytical} without this compensation.

Even in the most extreme cases, the magnitude of this offset is relatively small. For WASP-107b, for example, which has one of the highest atmospheric scale heights of any exoplanet, the offset is on the order of one second or less for modest ($N$ $<$ 2) limb asymmetry. Nevertheless, the existence of this extra-area compensation effect highlights the inherent challenge in accurately measuring a planet's time of conjunction. Not only must potential astrophysical biases, such as the timing bias or dynamical transit timing variations (discussed in the next section), be considered, but biases introduced by numerical techniques in the analysis can have an impact as well. Fortunately, this particular numerical bias is relatively small -- far smaller than the uncertainty of most current measurements achieved by Kepler, TESS, Spitzer, and ground-based observations -- and will only exist if the planet indeed has limb asymmetry. Still, this illustrates the types of timing biases that must be newly considered when performing high-precision transit observations with JWST.

    \subsection{Additional Effects on the Observability of Limb Asymmetry} \label{subsec:discussion_observability}

Some exoplanets exhibit transit timing variations (TTVs) induced by gravitational interaction with other planets in the same system \citep[e.g.][]{maciejewski2010, adams2011}. The amplitude of these TTVs can range from less than one second to tens of minutes, depending on the orbital architecture of the system, and have variation periods upward of hundreds of years. For certain planets, these TTVs may place a limit on the achievable precision on a planet's time of conjunction that is comparable or worse than the timing biases derived here. Therefore, unless the underlying TTV trend is known extremely well, it will be difficult to disentangle signatures of limb asymmetry from the uncertainty in the time of conjunction generated by TTV interactions for such planets.

Stellar surfaces are not perfectly homogeneous, and many exoplanet-hosting stars are known to have starspots and faculae on their surfaces. These stellar surface heterogeneities influence the signal observed during a planet's transit, which can lead to biased inferences of the planet's atmosphere if not properly accounted for \citep[e.g.][]{rackham2018TLSE1, rackham2019TLSE2, barclay2021tlsek218, moran2023tlsegj486}. We did not consider the impact of stellar surface heterogeneities in this work, and our results apply most directly to scenarios when any stellar contamination can be confidently ruled out or modeled. A more detailed study is warranted to examine how stellar contamination could further mimic or bias inferences of exoplanet limb asymmetry.

Finally, our analysis constructed the disk of an asymmetric-limb planet by explicitly assuming east--west asymmetry where each hemisphere is a uniform semicircle. Real exoplanet atmospheres likely do not follow this geometry. For example, strong meridional circulation that preferentially transports condensates to a planet's polar regions may cause pole--equator asymmetry that is more significant than east--west asymmetry \citep[e.g. see][]{parmentier2013_polarclouds, charnay2015_polarclouds, line2016}. In this case, the planetary disk would be shaped differently than we have assumed, either as a 90$^\circ$ rotation or even a non-semicircular shape, which may lead to a different form of the limb asymmetry - transit timing degeneracy than we have derived here. 

    \subsection{Potential Limitations of Our Analytic Model}\label{subsec:discussion_modellimitations}

Our analytic model neglects any rotation of the planetary disk during the transit. For a planet with asymmetric limbs, especially as constructed herein, the projected planetary disk will change slightly as the planet rotates slightly between ingress and egress. We neglected this rotation to keep the geometry as simple as possible, which will cause our formulae to slightly underestimate the timing bias -- for example, contributing to the outstanding residual in Figure~\ref{fig:perfectLC_timingbias}. Fortunately, for the planets we considered here, which have semi-major axes ranging $\sim$9 -- 18x their host stellar radii, this is a negligible effect. To test the maximum effect this will have, we repeated our calculations for the exoplanet WASP-12b, which has $a/R_\star = 3$ \citep{chakrabarty2019_wasp12b} and represents the population for which intratransit rotation is most important. In this case, our analytic model underestimated the timing bias by 0.39~s for $N$=1, 0.78~s for $N$=2, and 1.93~s for the extreme case of $N$=5. 

Similarly, in our analytic model, we assume the planets move in a flat plane along the stellar disk during transit. In reality, of course, orbits are curved so these paths are slightly nonlinear, especially for close-in orbits as above. To examine the validity of this assumption, we rederived the timing bias based on the transit duration derived by \citet{seager2003_orbits} which solves exactly for the sky-projected planet--star separation with time, and compared it to our calculations for WASP-39b and WASP-12b. The latter case, as before, represents the most close-in orbits possible where this approximation would break down. In both cases, though, the difference relative to our nominal value is effectively zero. 

As mentioned previously, the semicircle-based model that we use for our asymmetric-limb planetary disk, both when deriving our analytic model and when using \texttt{catwoman} in our numerical model, is not strictly realistic. As seen clearly in the exaggerated example in Figure~\ref{fig:exaggerated_intro_schematic}, this construction creates a discontinuity in the atmosphere near the planet's poles, which becomes more extreme for greater limb asymmetry. This limitation, however, is unlikely to be important given current and near-future data quality. For example, as seen in the residuals in Figure~\ref{fig:perfectLC_timingbias} and discussed in Section~\ref{subsec:numerical_extraareaproblem}, this discontinuity only induces error in our model of approximately one tenth of a second for limb asymmetry greater than five scale heights. While this was calculated just for one example exoplanet, the magnitude would not be significantly different for other planets. Transit timing at such high precision is likely not achievable with current instrumentation, as combined fits of state-of-the-art JWST, TESS, Spitzer, and ground-based data have only been able to achieve 1-$\sigma$ uncertainty on $t_c$ of $\sim$0.6~s or worse \citep{carter:inprep, murphy:inreview}. Also, it is not clear from current observations whether limb asymmetry as large as five scale heights and greater actually exists in nature. Nevertheless, analytic and numerical models that define fully continuous atmospheric surfaces, such as the transmission string numerical model \citep{grant2023_tstrings}, may improve upon our result.

\section{Summary} \label{sec:summary}

Measuring morning-to-evening variations of temperature, elemental and molecular abundances, and aerosol properties in exoplanet atmospheres is critical to understanding the underlying dynamics and chemistry of the atmosphere. Transit observations are able to extract the separate contributions from a planet's morning and evening terminator using high-resolution ground-based spectroscopy and, more recently, low-resolution space-based spectroscopy with JWST. A strong degeneracy exists between the effect of limb asymmetry on a light curve and uncertainty in the planet's time of conjunction. In this work, we discussed the origin of this degeneracy and derived a simple, physically motivated analytical model for it, highlighted by Equation~\ref{eqn:dt_analytic} and generalized in Equation~\ref{eqn:generaltimingbiaseqn}. We found that limb asymmetry of one to two scale heights difference in limb radii can cause the apparent time of conjunction to be biased by upward of 10 seconds. This bias is more extreme for planets with larger scale height, and as mentioned we derived a generalized formula (Equation~\ref{eqn:generaltimingbiaseqn}) for estimating this bias on any exoplanet.
When comparing our analytic model to simulated observations of an asymmetric-limb planet fit using a uniform-limb model, we found an additional numerical bias is added to to the inferred time of conjunction, which is typically less than one second. We dubbed this the ``extra-area compensation effect," as it stems from the relative occulting areas of a uniform-limb disk and asymmetric-limb disk when stacked atop one another. This effect is included in the above generalized formula, and we derived formulae (Equation~\ref{eqn:compensation_general_tau_lineend}, generalized in Equation~\ref{eqn:generalcompensationeqn}) specifically for this effect as well.

Our results underscore the importance of obtaining precise $t_c$ measurements of target exoplanets to be able to probe for limb asymmetry in their atmospheres. We have shown that times derived from single-filter observations can be strongly biased. Therefore, the recommended method to make precise $t_c$ measurements is to simultaneously fit transit observations, at the highest possible cadence and precision, from multiple wavelengths, as shown by \cite{powell2019}. Optical space-based observations (e.g., TESS, CHEOPS, and Kepler) are usually readily available, but have a lot of bandpass overlap. We would recommend combining these with nonoverlapping ground-based and/or near-infrared space-based observations (e.g., Spitzer/IRAC, and JWST) to get the best possible constraint. These observations need not be simultaneous with each other or the primary observation (i.e., the one being fitted for limb asymmetry), but will be most effective when as close in time to the primary observation as possible. 

\section*{Acknowledgements}
The authors would like to acknowledge Kevin Hardegree-Ullman, Megan Mansfield, and Everett Schlawin for helpful discussions that inspired this study. M. M. M. acknowledges funding from NASA Goddard Spaceflight Center via NASA contract NAS5-02105. This material is based upon work supported by the National Aeronautics and Space Administration under Agreement No. 80NSSC21K0593 for the program “Alien Earths”. The results reported herein benefited from collaborations and/or information exchange within NASA’s Nexus for Exoplanet System Science (NExSS) research coordination network sponsored by NASA’s Science Mission Directorate. This research has made use of NASA’s Astrophysics Data System Bibliographic Services. This research has made use of the NASA Exoplanet Archive, which is operated by the California Institute of Technology, under contract with the National Aeronautics and Space Administration under the Exoplanet Exploration Program.

\vspace{5mm}
\facilities{Exoplanet Archive}
\software{
        batman \citep{batman},
        catwoman \citep{Espinoza2021_catwoman2, catwoman1},
        emcee \citep{emcee},
        matplotlib \citep{matplotlib},
        NumPy \citep{numpy},
        PandExo \citep{pandexo},
        SciPy \citep{scipy}
          }






\bibliography{refs}{}
\bibliographystyle{aasjournal}



\end{document}